\documentclass[3p,times]{elsarticle}

\usepackage{ecrc1}

\newcommand{\beq}[1]{\begin{equation}\label{#1}}
\newcommand{\eeq}{\end{equation}}
\newcommand{\bea}[1]{\begin{eqnarray}\label{#1}}
\newcommand{\eea}{\end{eqnarray}}

\def\ktr{\tilde{\kappa}_{\rm tr}}
\def\etal{{\it et al.}}

\def\lsim{\mathrel{\rlap{\lower3pt\hbox{$\sim$}}
    \raise2pt\hbox{$<$}}}
\def\gsim{\mathrel{\rlap{\lower3pt\hbox{$\sim$}}
    \raise2pt\hbox{$>$}}}

\def\half{{\textstyle{1\over 2}}}

\newcommand{\rf}[1]{(\ref{#1})}

\volume{17}

\firstpage{135}

\journalname{Physics Procedia}

\runauth{Ralf Lehnert}

\jid{phpro}

\jnltitlelogo{Physics Procedia}

\usepackage{amssymb}

\usepackage[figuresright]{rotating}

\begin{document}

\begin{frontmatter}



\dochead{\large Physics of Fundamental Symmetries and Interactions -- PSI2010}

\title{Spacetime-symmetry violations: motivations, phenomenology, and tests}


\author{Ralf Lehnert}

\address{Instituto de Ciencias Nucleares,
Universidad Nacional Aut\'onoma de M\'exico,
Apartado Postal 70-543, M\'exico, 04510, D.F., Mexico}
\ead{ralf.lehnert@nucleares.unam.mx}

\begin{abstract}
An important open question in fundamental physics concerns the nature of spacetime at distance scales associated with the Planck length. The widespread belief that probing such distances necessitates Planck-energy particles has impeded phenomenological and experimental research in this context. However, it has been realized that various theoretical approaches to underlying physics can accommodate Planck-scale violations of spacetime symmetries. This talk surveys the motivations for spacetime-symmetry research, the SME test framework, and experimental efforts in this field.
\end{abstract}


\end{frontmatter}


\section{Introduction}
\label{intro}

Spacetime plays a fundamental role in science: 
it not only provides the arena in which physical processes take place, 
but it also exhibits its own dynamics. 
Like many other basic physical entities, 
spacetime is, 
at least partially, 
characterized by its underlying symmetries. 
The continuous ones of these symmetries comprise four spacetime translations 
and six Lorentz transformations 
(three rotations and three boosts), 
which are intertwined in the Poincar\'e group.
Because of its fundamental importance, 
various aspects of Poincar\'e invariance
have been tested in the past century 
with no credible experimental evidence for deviations
from this symmetry. 
It is fair to say that 
Poincar\'e invariance
(and in particular Lorentz symmetry)
has acquired a venerable status in physics.

Nevertheless, 
the last decade has witnessed a renewed interest 
in spacetime-symmetry physics 
for various reasons. 
First, 
the quantum structure of spacetime 
is likely to be no longer a smooth four-dimensional manifold. 
This suggests that 
Planck-suppressed deviations from 
the usual four-dimensional classical Poincar\'e symmetry 
could be a promising quantum-gravity signature. 
Indeed, 
many leading theoretical approaches to unify quantum physics and gravity, 
such as 
string theory~\cite{kps},
spacetime foam~\cite{sf},
non-commutative field theories~\cite{nc},
and cosmological supergravity models~\cite{varscal}, 
can accommodate minute departures from relativity theory.
Second, 
varying couplings 
(which break translation symmetry)
driven by scalar fields 
are natural in many theoretical models 
beyond established physics~\cite{DamourPolyakov94}.
Moreover, 
there have been recent experimental claims 
of a spacetime-dependent fine-structure parameter $\alpha$~\cite{varyingexp}.

To identify and analyze suitable experiments that 
can provide ultra-sensitive tests of Lorentz invariance, 
an effective field theory called the Standard-Model Extension (SME)
has been developed~\cite{sme}. 
The SME essentially contains the entire body of established physics
in the form of the Lagrangians for the usual Standard Model and general relativity.
This fact guarantees that
practically all physical systems
can be investigated with regards to their potential to test Lorentz symmetry. 
The description of Lorentz violation is achieved with additional lagrangian terms that 
are formed by covariant contraction of conventional fields 
with background vectors or tensors 
assumed to arise from underlying physics. 
These nondynamical vectors or tensors 
represent the SME coefficients controlling the nature and size 
of potential violations of Lorentz symmetry.
The set of all such correction terms 
leads to the full SME~\cite{nonren},
whereas the subset of relevant and marginal operators 
results in the minimal SME (mSME).

Most recent experimental investigations of Lorentz invariance 
have been performed within the mSME.
Specific studies include, 
for instance, 
ones with photons~\cite{randomphotonexpt,cherenkov}, 
neutrinos~\cite{randomnuexpt},
electrons~\cite{randomeexpt}, 
protons and neutrons~\cite{randompnexpt}, 
mesons~\cite{randomhadronexpt}, 
muons~\cite{muexpt},
and gravity~\cite{gravity}.  
Several of the obtained experimental limits 
exhibit sensitivities that
can be regarded as testing Planck-scale physics. 
A tabulated overview of tests and their results 
can be found in Ref.~\cite{tables}.
No solid experimental evidence for deviations from Lorentz symmetry 
has been found to date, 
but discovery potential might exist in neutrino physics~\cite{neutrinomodel}.

The outline of this presentation is as follows.
In Sec.~\ref{interplay}, 
various spacetime symmetries are reviewed 
with particular focus on their interplay. 
Section~\ref{smesec}
presents the basic ideas and the philosophy
that underpin the construction of the SME.
Three examples for mechanisms that
can generate Lorentz-invariance breakdown 
in Lorentz-symmetric underlying models 
are contained in Sec.~\ref{mechanisms}. 
Section \ref{tests} 
describes some experimental Lorentz tests 
in different physical systems. 
A brief summary 
is given in Sec.~\ref{sum}.

\section{Spacetime symmetries and relations between them}
\label{interplay}

In conventional nongravitational physics, 
the four spacetime translations 
(i.e., three spatial translations and one time translation)
form an exact symmetry 
and therefore lead to the conservation of 4-momentum.
One mathematical condition for translation invariance 
is the absence of explicit spacetime dependencies in the Lagrangian.
But it is known that 
various theoretical approaches to physics beyond the Standard Model 
can lead to varying couplings.
It follows that 
in such approaches
translation symmetry is typically be violated. 
This idea is presently also attracting attention 
because of observational claims of a varying $\alpha$,
as mentioned in the introduction.

Suppose translation symmetry is indeed broken.
It is then natural to ask 
whether other symmetries, 
and in particular Lorentz invariance,
can be affected.
To answer this question, 
we start by looking at the generator for Lorentz transformations, 
which is the angular-momentum tensor $J^{\mu\nu}$:
\begin{equation} 
J^{\mu\nu}=\int d^3x \;\big(\Theta^{0\mu}x^{\nu}-\Theta^{0\nu}x^{\mu}\big)\;. 
\label{gen} 
\end{equation}
Here, 
$\Theta^{\mu\nu}$ denotes the energy--momentum tensor, 
which is associated with spacetime translations.
Since translation do not represent a symmetry 
in the present context, 
$\Theta^{\mu\nu}$ is now no longer conserved.
As a consequence, 
$J^{\mu\nu}$ 
will typically become spacetime dependent 
in models with varying couplings.
In particular, 
the usual spacetime-independent 
Lorentz-transformation generators cease to exist. 
It follows that 
exact Lorentz invariance 
is not guaranteed. 
We conclude that 
(with the exception of special circumstances) 
translation-symmetry violation
leads to Lorentz-invariance breakdown. 

Let us continue along this hypothetical avenue 
and suppose Lorentz symmetry is broken.
Together with locality, 
quantum mechanics,
and a few other mild conditions, 
Lorentz symmetry 
is a key ingredient for the celebrated 
CPT theorem discovered by Bell, L\"uders, and Pauli 
over half a century ago. 
Now, 
the question arises 
whether the absence of Lorentz symmetry 
would affect CPT invariance.
Unfortunately, 
there is no clear-cut answer to this question.
For example, 
in the SME to be discussed in the next section,
about half of the relevant and marginal operators
for Lorentz breaking 
also violate CPT symmetry.
However, 
we may also consider a slightly different question 
and ask which one of the ingredients for the CPT
theorem should be dropped, 
if we want to investigate CPT violation.
Clear is that
not all of the assumptions for the CPT theorem 
can survive simultaneously
because this would exclude CPT breaking.

The answer to this question
is largely dependent on 
the physics
causing CPT breakdown. 
A general and mild assumption is that
the low-energy leading-order effects of new physics 
are describable 
by a local effective field theory:  
such theories represent an immensely flexible framework,
and they have been successful in various subfields of physics 
including solid-state, nuclear, and particle physics.
In such a context,
it appears unavoidable 
that the property of exact Lorentz invariance needs to be relaxed.
This expectation can be proven rigorously 
in axiomatic quantum field theory~\cite{green02,green}.
This result, 
sometimes called ``anti-CPT theorem,'' 
roughly states that
in any unitary, local, relativistic point-particle field theory 
CPT breakdown comes with Lorentz violation.
However, 
as we have noted above,
the converse of this statement
(i.e., Lorentz breaking implies the loss of CPT symmetry) 
is false in general. 
It is thus apparent 
that under the above general and plausible assumption,
CPT tests also probe Lorentz invariance.
We note 
that other types of CPT breaking 
resulting from apparently non-unitary quantum mechanics
have also been discussed in the literature~\cite{mav}.

\section{Building the SME}
\label{smesec}

To study the low-energy effects of Lorentz and CPT violation,
both theoretically and phenomenologically,
a comprehensive test framework is needed.
Early test models for special relativity 
were seeking to parametrize deviations 
from the Lorentz transformations. 
Examples of models of this type are
Robertson's framework~\cite{Robertson49}, 
its Mansouri--Sexl extension~\cite{MSnodel}, 
and the $c^2$ model~\cite{csquared}.
More recently, 
also other, quantum-gravity motivated approaches, 
such as phenomenologically constructed modified one-particle dispersion relations, 
have been considered.
These models have in common that 
they focus solely on kinematical deviations from Lorentz symmetry.
This often provides the advantage of conceptual simplicity 
when applied to experimental situations.
On the other hand,
their behavior under the CPT transformation is typically unclear. 
Moreover, 
the absence of dynamics implies that 
only a limited range of tests can be identified and analyzed.
The SME, 
already mentioned in the introduction,
has been developed to avoid these issues.
This section describes the cornerstones
on which the SME test framework has been constructed. 

We begin by reasoning in support of a test model 
that describes dynamical in addition to kinematical physics properties.
It is true that
a certain set of kinematical laws may be compatible 
with various dynamical models 
suggesting a larger degree of generality.
Nevertheless, 
the dynamics of realistic models is restricted by the condition that
established physics must emerge 
under certain conditions. 
In addition, 
it appears difficult if not impossible 
to create a framework for deviations from Lorentz symmetry 
that contains the usual Standard Model
while at the same time possessing dynamics that
is substantially different from that of the SME 
to be discussed below.
Finally, 
most potential signals for Lorentz and CPT violation 
involve some kind of dynamics, 
or are incomplete without additional dynamical checks, 
as mentioned above.
For these reasons, 
it seems advantageous
to have at one's disposal 
a fully dynamical test framework 
for Lorentz and CPT violation.  

{\bf Construction of the SME.}
To appreciate 
the generality of the SME, 
we briefly describe the philosophy and main ideas 
behind its construction~\cite{sme,nonren}.
The starting point for establishing the SME
is the entire body of known physics 
in the form of the conventional Standard-Model and Einstein--Hilbert Lagrangians 
${\mathcal L}_{\rm SM}$ and ${\mathcal L}_{\rm EH}$, respectively.
To implement Lorentz and CPT violation, 
the most general set of lagrangian correction terms $\delta {\mathcal L}_{\rm LV}$ 
(compatible with otherwise desirable feaures) 
is then included: 
\begin{equation} 
{\mathcal L}_{\rm SME}=\underbrace{{\mathcal L}_{\rm SM}+{\mathcal L}_{\rm EH}}
_{\textrm{established physics}}
+\underbrace{\delta {\mathcal L}_{\rm LV}}_{\textrm{Lorentz violation}}\; . 
\label{sme} 
\end{equation}
In the above equation, 
the SME Lagrangian is denoted by ${\mathcal L}_{\rm SME}$. 
The terms $\delta {\mathcal L}_{\rm LV}$ 
describing the nature and extent of Lorentz and CPT breakdown
can in principle be of any mass dimensionality.
They are built by covariant contraction of Standard-Model and gravitational fields  
with Lorentz-breaking tensorial coefficients 
yielding scalars, 
which ensures coordinate independence. 
The nondynamical tensorial coefficients  
represent a nontrivial vacuum 
with background vectors or tensors, 
which violate Lorentz symmetry 
and in some cases also CPT invariance.
These background vectors and tensors 
are assumed to be generated 
by more fundamental physics,
such as quantum-gravity models.
It then becomes apparent 
that the entire set  of possible contributions to $\delta {\mathcal L}_{\rm LV}$ 
yields the most general effective field theory 
of leading-order Lorentz violation 
at the level of an observer Lorentz-invariant 
unitary Lagrangian.

Examples of terms contained in the flat-spacetime limit of the SME 
are the following:
\begin{equation}
\label{sampleterms}
\delta {\mathcal L}_{\rm LV}\supset 
b^{\mu}\overline{\psi}\gamma_5\gamma_{\mu}\psi\,,\;
(r^{\mu}\overline{\psi}\gamma_5\gamma_{\mu}\psi)^2\,,\;
(k_F)^{\alpha\beta\gamma\delta}F_{\alpha\beta}F_{\gamma\delta}\,,\;
\half\,\epsilon_{\alpha\beta\gamma\delta}\,(k_{AF})^\alpha A^\beta F^{\gamma\delta}\,.
\end{equation}
In these expressions,
$\psi$, $F$, and $A$ denote a conventional spinor field
and a conventional gauge field strength, 
and a conventional gauge potential
respectively.
The quantities $b^{\mu}$, $r^{\mu}$, and $(k_F)^{\alpha\beta\gamma\delta}$ 
are SME coefficients controlling the size and type of Lorentz violation. 
They are taken as caused by underlying physics, 
perhaps by some quantum-gravity model.
An experiment would seek to measure or constrain these coefficients.
The mSME mentioned in the introduction
is restricted by further physical requirements, 
such as translational invariance, 
and power-counting renormalizability. 
For instance, 
the mSME does not contain the $r^{\mu}$ term 
shown in the above expression~\rf{sampleterms}.

We note in passing
that in a curved-manifold context involving gravitational physics, 
this idea is most easily implemented 
utilizing the vierbein formalism.
A key result in this context is that 
explicit Lorentz violation typically results in an incompatibility 
between the Bianchi identities 
and the covariant conservation laws for the energy--momentum and spin-density tensors.
However, 
a spontaneous breakdown of Lorentz invariance 
avoids this issue, 
so that some type of dynamical symmetry violation 
is favored for generating SME coefficients. 
Examples of such mechanisms 
are given in the next section. 
Two consequences for the present discussion are that
when gravity cannot be neglected,
SME coefficients would need to exhibit 
a certain spacetime dependence as dictated by compatibility, 
and new degrees of freedom may occur.

The SME is flexible enough 
to incorporate additional potential features of new physics, 
such as non-pointlike elementary excitations 
or a fundamental discreteness of spacetime at the Planck scale.
It is therefore improbable that 
the above effective-field-theory approach 
is insufficient 
at currently attainable energies. 
One may even argue 
that presently established physics 
(i.e., the Standard Model and general relativity)
is also regarded as a low-energy effective description 
of underlying physics.
It would then seem surprising 
if potential Lorentz-violating effects from such underlying physics
could not also be described within effective field theory.
We finally remark 
that the requirement for a low-energy description 
{\it outside} the framework of effective field theory 
is difficult to imagine 
for new physics 
with novel Lorentz-{\it invariant} features, 
such as further particle species or scalar fields, 
new symmetries, 
or additional spacetime dimensions. 
Notice in particular 
that Lorentz-symmetric modifications 
can therefore easily be included into the SME, 
if it becomes necessary~\cite{susy}. 

{\bf Benefits of the SME.}
The SME 
permits the identification,
analysis, 
and direct comparison 
of practically all currently feasible experiments 
that can search for deviations from Lorentz and CPT invariance. 
In certain limiting cases of the SME, 
one can recover classical kinematics test models of special relativity 
(such as the aforementioned framework by Robertson, 
its Mansouri--Sexl extension to arbitrary clock synchronizations, 
or the $c^2$ model)~\cite{nonren,km02}. 
An additional advantage of the SME 
is the flexibility of including
further desirable features 
besides coordinate independence. 
For instance, 
it is possible to impose 
spacetime-translation invariance 
(at least in the flat-spacetime limit), 
SU(3)$\times$SU(2)$\times$U(1) gauge invariance, 
power-counting renormalizability, 
unitarity,
and locality. 
These additional features
put further constraints on the parameter space for Lorentz and CPT violation. 
Another possibility is 
to make simplifying choices, 
such as a residual rotational invariance
in certain classes of inertial coordinate systems. 
For example, 
the latter hypothesis 
together with additional simplifications of the SME 
has been considered in some investigations~\cite{cg99}. 

{\bf Consistency of the SME.}
Thus far,
we have reviewed just the general idea for building the SME framework. 
One may also inquire about its theoretical consistency. 
To date, 
there have been a number of more formal and theoretical investigations
within the SME.
They have addressed subjects such as
radiative corrections~\cite{radcorr},
renormalizability~\cite{ren},
supersymmetry~\cite{susy},
causality~\cite{caus},
kinematics~\cite{disprel},
symmetry studies~\cite{symstud},
higher-derivative terms~\cite{higherder},
gravity~\cite{grav},
and mathematical studies~\cite{math}.
None of these investigations 
has found inconsistencies 
or other difficulties that 
would render the SME 
unsuitable for describing Lorentz and CPT violation.

\section{Generating Lorentz breakdown}
\label{mechanisms}

Thus far,
we have examined how
the breakdown of one spacetime symmetry
can also lead to the violation of another spacetime invariance,
and we have constructed by hand
a low-energy effective description for such effects.
Another key question concerns actual mechanisms 
within theoretical approaches to physics beyond the Standard Model
that can lead to symmetry breaking in the first place.
In the present section,
we set out to address this question
by providing some intuition 
regarding possible sources for Lorentz violation
in candidate fundamental models. 
A number of possible mechanisms
have already been mentioned in the introduction.
Here, 
we will discuss three of them---spontaneous Lorentz breakdown, 
Lorentz violation through varying couplings, 
and non-commutative field theory---in some more detail.

{\bf Spontaneous Lorentz and CPT violation.} 
The mechanism of spontaneous symmetry breakdown
is well established in various subfields of physics.
For example,
it can occur in the physics of elastic media
and in condensed-matter physics.
This mechanism is also part of the Standard Model of particle physics.
From a theoretical point of view, 
this mechanism is quite appealing 
because of the following.
In many circumstances, 
the internal consistency of a QFT requires the presence of a symmetry.
However, the symmetry is not observed in nature.
Spontaneous symmetry violation resolves such situations:
The dynamical underpinnings of the model 
remain symmetric,
which ensures consistency.
On the other hand, 
the ground-state solution 
(which essentially corresponds to the observed physical system)
fails to exhibit the full symmetry of the model.

The key ingredient for spontaneous symmetry breakdown 
is an interaction that 
destabilizes the naive vacuum 
and triggers a vacuum expectation value.
This can, for instance, 
be achieved with a potential-energy term in the Lagrangian.
As an example
consider a Higgs-type field $\varphi$
whose expression for the potential-energy density
is given by $V(\varphi)=g (\varphi^2-\lambda^2)^2$.
Here, 
$\lambda$ and $g$ are constants. 
We note in passing that 
a possible spacetime dependence  $\varphi=\varphi(x)$ 
would result in additional, positive-valued contributions
to the energy density, 
which permits us to focus solely on a constant  $\varphi$.
The basic idea now is that 
the vacuum is usually taken as the state with the lowest energy. 
The lowest-energy configuration {\it requires} 
$\varphi$ to be nonzero: $\varphi=\pm\lambda$. 
As a consequence, 
the physical vacuum 
for a system involving a Higgs-type field $\varphi$
is not empty; 
it contains, in fact, 
the condensate of the spacetime-constant scalar field 
$\varphi_{\rm vac}\equiv\langle\varphi\rangle=\pm\lambda \neq 0$,
where $\langle\varphi\rangle$ 
denotes the vacuum expectation value (VEV) of $\varphi$. 
It is important to notice
that $\langle\varphi\rangle$ is a Lorentz scalar, 
and thus it does {\it not} select a preferred direction in spacetime 
leaving Lorentz symmetry intact. 

This situation changes 
when the scalar field $\varphi$ 
is replaced by a vector or tensor field.
For simplicity,
let us consider a 3-vector field $\vec{R}$ as an example~\cite{culiacan}.
The relativistic generalization to 4-vectors or 4-tensors is
relatively simple. 
Neither the $\vec{R}$ field 
nor its relativistic generalizations
are present in the Standard Model, 
and there is currently no experimental evidence for such types of field. 
Nevertheless, 
additional vector fields like $\vec{R}$ 
are contained in numerous candidate fundamental theories. 
Paralleling the previous Higgs-type case, 
we posit the following the expression for the energy density of $\vec{R}=\textrm{const.}$:
\begin{equation} 
\label{vec_en_den} 
V(\vec{R})=(\vec{R}^2-\lambda^2)^2\, . 
\end{equation} 
We see that 
the lowest possible energy 
associated with the vacuum state
is zero.
As in the previous $\varphi$-field example, 
this lowest energy requires $\vec{R}$ to be nonzero:
$\vec{R}_{\rm vac}\equiv\langle\vec{R}\rangle=\vec{\lambda}$, 
where $\vec{\lambda}$ is any constant vector satisfying $\vec{\lambda}^2=\lambda^2$. 
Again, 
the vacuum does not stay empty; 
it actually contains the VEV of the vector field, $\langle\vec{R}\rangle$. 
It follows that
the true vacuum in the above model 
possesses an intrinsic direction 
given by $\langle\vec{R}\rangle$. 
The point is 
that such an intrinsic direction
violates rotation symmetry and therefore Lorentz invariance. 
We remark 
that interactions generating energy densities like those in Eq.~\rf{vec_en_den} 
are absent in conventional renormalizable gauge theories, 
but they may occur in the context of string field theory, 
for example. 

{\bf Spacetime-dependent scalars.} 
A spacetime-dependent scalar, 
such as a cosmologically varying coupling
typically leads to the breakdown of spacetime-translation invariance \cite{varscal},
regardless of the mechanism causing this dependence.
Since a fundamentally varying scalar violates translation invariance,
it will typically also break Lorentz symmetry,
as was explained in Sec.~\ref{interplay}. 
Here, 
we will focus on an explicit example for this effect. 

Consider a physical system with a varying coupling denoted by $\xi(x)$ 
and two scalar fields $\phi$ and $\Phi$. 
Suppose further that the Lagrangian $\mathcal{L}$ 
contains a kinetic-type interaction of the form 
$\xi(x)\,\partial^{\mu}\phi\,\partial_{\mu}\Phi$. 
Under mild assumptions, 
one may integrate by parts the action associated with this Lagrangian
(for example with respect to the first partial derivative in the above term) 
without any change in the equations of motion. 
An equivalent Lagrangian $\mathcal{L}'$ is then given by
\begin{equation}
\mathcal{L}'\supset -K^{\mu}\phi\,\partial_{\mu}\Phi\, .
\label{example1}
\end{equation}
Here, $K^{\mu}(x)\equiv\partial^{\mu}\xi(x)$ is an external
nondynamical 4-vector. 
It is apparent that 
this 4-vector selects a preferred direction in spacetime, 
which violates Lorentz invariance. 
We note
that for variations of $\xi(x)$ on cosmological scales, 
$K^{\mu}$ is approximately spacetime constant locally 
(e.g., on solar-system scales) 
to an excellent approximation. 

\begin{figure}[h]
\begin{center}
\includegraphics[width=0.50\hsize]{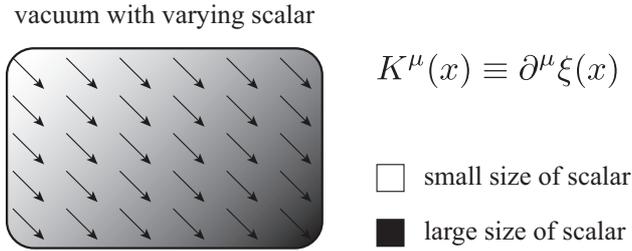}
\caption{Lorentz-invariance violation via varying scalars.
The background shading of gray represents the magnitude of the scalar: 
the darker regions correspond to larger values of the scalar. 
The black arrows represent the gradient $K^{\mu}(x)\equiv\partial^{\mu}\xi(x)$ of the scalar, 
which determines a preferred direction in spacetime.
It follows that Lorentz invariance is violated. 
}
\label{fig1} 
\end{center}
\end{figure} 

Intuitively, 
the violation of Lorentz invariance 
as a result of a varying scalar can be visualized as follows. 
The 4-gradient of the varying scalar must clearly be nonzero---at least 
in some spacetime region. 
Otherwise,
the scalar would be constant. 
This 4-gradient 
then picks out a preferred direction 
in this region, 
as is illustrated in Fig.~\ref{fig1}. 
Consider, 
for instance, 
a particle 
that exhibits certain interactions with the varying scalar. 
Its propagation properties 
might be affected differently
in the directions parallel and perpendicular to the gradient. 
But directions that are physically inequivalent 
must lead to rotation-symmetry breaking. 
Because rotations are contained in the Lorentz group, 
Lorentz symmetry must be broken. 

{\bf Non-commutative field theory.} 
An approach to quantum gravity that
has been gaining popularity for some time now 
is non-commutative field theory.
Roughly speaking, 
the basic idea is to achieve some description of a quantum spacetime 
by promoting coordinates to operators: $x^\mu\to\hat{x}^\mu$.
As a result, 
the $\hat{x}^\mu$ no longer commute;
they obey the relation
\begin{equation}
\label{noncom}
\left[\hat{x}^\mu,\hat{x}^\nu\right]=i\theta^{\mu\nu}\,.
\end{equation}
Here, 
$\theta^{\mu\nu}$ is a spacetime constant tensor.\footnote{Contrary to our earlier remarks in Sec.~\ref{smesec}, the Lorentz violation is explicit at this level. 
However, non-commutativity with the effective description~\rf{noncom} 
can also be generated dynamically in underlying physics.}
As such, 
$\theta^{\mu\nu}$ 
selects preferred directions in spacetime
violating Lorentz symmetry.
To see explicitly this violation,
the theory must be interpreted properly.
It turns out that
in some circumstances a model on a non-commutative spacetime
can be mapped to a quantum field theory 
on conventional Minkowski space.
This procedure generates lagrangian terms 
like $\theta^{\alpha\beta}F_{\alpha\mu}F_{\beta\nu}F^{\mu\nu}$,
which are present in the SME~\cite{nc}.
It thus becomes apparent that
non-commutative field theories provide a third mechanism that
can lead to Lorentz breakdown 
describable within the SME.

\section{Testing Lorentz violation}
\label{tests}

Because operators of all mass dimensionalities are allowed in the full SME, 
an infinite number of Lorentz- and CPT-breaking coefficients arise.
However, 
in an effective quantum field theory 
one might generically expect the relevant and marginal operators 
to dominate in the low-energy limit.
As mentioned in the introduction, 
the restriction to this subset of SME operators
is referred to as the mSME.
This section contains a brief overview 
of a representative sample of experimental efforts 
that are primarily concerned with mSME coefficients.

{\bf Kinematical tests with particle collisions.}
A widely known effect of Lorentz breakdown is
the modification of one-particle dispersion relations.
Such modified dispersion relations 
would generate corrections in the energy--momentum conservation equations
in particle collisions. 
We remark in passing that
the flat-spacetime mSME coefficients are taken as spacetime constant,
so translational symmetry, 
and thus energy--momentum conservation, 
still hold.
The Lorentz-violating corrections to the collision kinematics could,
for example, 
cause the following effects:
reaction thresholds may be shifted,
reactions kinematically forbidden in Lorentz-symmetric physics
may now occur,
and certain conventional reactions 
may no longer be allowed kinematically.

Consider, 
for instance,
the spontaneous emission of a photon 
from a free charge.
In ordinary physics, 
the conservation of energy and momentum 
does not permit this process to occur.
Nevertheless, 
certain types of Lorentz and CPT violation
can cause the speed of light to be slower 
relative to the speed of, say, electrons.
In analogy to conventional Cherenkov radiation
(when light propagates slower inside a macroscopic medium with refractive index $n>1$),
free electrons can now emit Cherenkov photons 
in a Lorentz- and CPT-breaking vacuum~\cite{cg99,cherenkov}.
This so-called ``vacuum Cherenkov effect'' 
may or may not exhibit a threshold
depending on the type of mSME coefficient.
In what follows, 
we consider mSME coefficients 
that lead to a threshold 
for vacuum Cherenkov radiation.
In this case,
one can extract an observational limit on the size of these mSME coefficients, 
as we will describe next.
Electrons propagating faster 
than the modified speed of light 
would slow down and fall below threshold 
through energy loss 
due to the emission of vacuum Cherenkov radiation. 
We can then conclude that
if highly energetic stable electrons exist in nature,
they cannot be above threshold. 
This information 
gives a lower bound for the threshold,
which in turn yields a limit on Lorentz violation.
Using, 
in this context, 
data from LEP electrons with energies up to $104.5\,$GeV
determines the constraint $\ktr-\frac{4}{3}c^{00}\lsim1.2\times10^{-11}$~\cite{collider}.
Here, 
$\ktr$ and $c^{00}$ parametrize the isotropic, polarization-independent, 
mass-dimension four SME operators for the photon and the electron, 
respectively.

We next consider the decay of photons in vacuum.
This is another particle reaction process 
not allowed kinematically by energy--momentum conservation
in conventional physics.
However, 
certain combinations of mSME coefficients 
can cause light to travel faster than 
the maximal attainable speed of electrons. 
In analogy to the vacuum-Cherenkov case discussed above,
in which high-energy electrons become unstable, 
we expect now that 
high-energy photons can become unstable against decay 
into an electron--positron pair.
With the modified dispersion relations 
emerging from the mSME,
one can indeed confirm 
that this expectation is met.
As in the case of the vacuum Cherenkov effect,
photon decay in a Lorentz-breaking vacuum 
often occurs above a threshold, 
and it can then be used to determine
an observational constraint 
on this particular type of Lorentz violation. 
The reasoning is as follows.
Suppose high-energy stable photons are observed. 
They must then essentially be below the threshold for photon decay.
This implies that 
the threshold energy must lie above 
the energy of these stable photons.
This bound on the threshold energy 
can be converted into a constraint on the size 
of the corresponding type of Lorentz breakdown.
Stable photons with energies up to $300\,$GeV
were observed at the Tevatron.
In this case, 
our reasoning yields the limit 
$-5.8\times10^{-12}\lsim\ktr-\frac{4}{3}c^{00}$~\cite{collider}.

We remark that the above bounds assume that
the rates for both vacuum Cherenkov radiation and photon decay 
are sufficiently fast.  
The purely kinematical reasoning 
we have presented above
is by itself inadequate 
for conservative experimental constraints.
This goes hand in hand 
with the discussion in Sec.~\ref{smesec} 
arguing that 
a dynamical framework is desirable,
and the full mSME 
(not only the predicted modified dispersion relations)
are required.
Appropriate studies within the mSME
indeed show that
the rates for vacuum Cherenkov radiation 
and photon decay 
would be efficient enough 
to validate the above arguments~\cite{BA08,ks08,collider}.

{\bf Spectropolarimetry of cosmological sources.}
There is one vectorial coefficient in the mSME's 
photon sector 
that breaks both Lorentz and CPT symmetry.
The corresponding operator is of mass dimension three,
and its structure is that of a Chern--Simons interaction.
It is parametrized by a background 4-vector usually denoted $(k_{AF})^\mu$ . 
Among the physical effects caused by the $(k_{AF})$ operator
is birefringence of electromagnetic waves~\cite{kAF}, 
vacuum Cherenkov radiations~\cite{cherenkov}, 
as well as certain frequency shifts in cavities~\cite{cavity}. 
These effects absent in known physics
are amenable to experimental inquiries.
Birefringence studies of electromagnetic radiation from cosmological sources 
are particularly well suited to search for this term:
the extremely long propagation distance
directly converts into ultrahigh sensitivity 
to this type of Lorentz and CPT violation.
Spectropolarimetric investigations of astrophysical data 
have established an upper bound on $(k_{AF})^\mu$ 
at the level of $10^{-42}\ldots10^{-43}\,$GeV~\cite{nonren,kAF}.

{\bf Spectroscopy of cold antihydrogen.} 
Comparative spectroscopy of hydrogen (H) and antihydrogen ($\overline{\rm H}$) 
is an excellent test of Lorentz and CPT breakdown. 
There are various transitions 
that can be studied. 
One of these is the unmixed 1S--2S transition, 
which appears to be an attractive candidate: 
the projected experimental sensitivity for this line 
is expected to be approximately at the level of $10^{-18}$.
This sensitivity  is promising in light of the anticipated Planck-scale suppression 
of quantum-gravity effects. 
However, 
an mSME study at first order in Lorentz and CPT violation
predicts the same shifts for free H and $\overline{\rm H}$ 
in the initial and final levels 
with respect to the usual energy states. 
From this perspective, 
the 1S--2S transition is actually less useful 
for the determination of unsuppressed Lorentz- and CPT-violating effects.
Within the mSME,
the leading non-trivial correction to this transition 
is generated by relativistic effects, 
and it enters with two further powers 
of the fine-structure parameter $\alpha$. 
The predicted modifications in the transition energy, 
already expected to be minuscule at zeroth order in $\alpha$, 
come therefore with a further suppression factor 
of more than ten thousand~\cite{antiH}. 

Another spectral line that
can be utilized for Lorentz and CPT tests 
is the spin-mixed 1S--2S transition.
When H or $\overline{\rm H}$ is confined in an electromagnetic trap,
such as a Ioffe--Pritchard trap,
the 1S and the 2S levels are each split 
as a consequence of the usual Zeeman effect. 
An mSME calculation for this case then shows that
the 1S--2S transition between the spin-mixed levels 
is indeed affected by Lorentz and CPT breakdown 
at leading order. 
A drawback from a practical viewpoint 
is the dependence of this transition on the magnetic field inside the trap,
so that the experimental sensitivity is limited 
by the size of the inhomogeneity of the trapping field $\vec{B}$. 
The development of novel experimental techniques 
might circumvent this problem, 
and a frequency resolutions close to the natural linewidth
might then be achievable~\cite{antiH}. 

A third transition that
is attractive for Lorentz- and CPT-violation studies
is the hyperfine Zeeman transition within the 1S state itself.
Even in the limit of a vanishing $\vec{B}$ field, 
mSME calculations show that  
there are first-order level shifts 
in two of the transitions
between the Zeeman-split states.
We note 
that this result may also be beneficial 
from an experimental point of view
because a variety of other transitions of this type, 
like the conventional H-maser line,
can be well resolved in the laboratory~\cite{antiH}. 

{\bf Experiments in Penning traps.}
The mSME predicts not only that
atomic energy levels can be shifted 
by the presence of Lorentz and CPT violation, 
but also, for example, 
the levels of protons and antiprotons 
inside a Penning trap. 
A perturbative calculation shows 
that only a single mSME coefficient 
(a CPT-violating $b^\mu$-type background vector, 
which is coupled to the chiral current of a fermion)
affects the transition-frequency shifts in the proton case
differently from those in the antiproton case
at leading order.
To be more specific, 
the anomaly frequencies are shifted in opposite directions 
for protons and their antiparticles. 
This effect can be utilized to 
extract a clean experimental constraint 
on the proton's $b^\mu$ coefficient~\cite{randomeexpt}.

{\bf Neutral-meson interferometry.} 
A well established and widely known CPT-invariance test 
compares the K-meson's mass 
to that of the corresponding antimeson:
even tiny mass differences 
would give measurable effects in Kaon-interferometry experiments.
In spite of the fact that 
the mSME contains only a single mass operator 
for a given quark--antiquark species, 
particle and antiparticle are nevertheless affected differently 
by the Lorentz- and CPT-violating background in the mSME. 
This generates different dispersion relations for a meson and its antimeson, 
so that mesons and antimesons can exhibit distinct energies 
at equal 3-momenta. 
It is this energy split that 
would ultimately be responsible for interferometric signals,
and it is therefore potentially observable in meson oscillations~\cite{randomhadronexpt}. 
We note 
that not only the K-meson 
but also other neutral mesons can be studied.
Notice in particular 
that in addition to CPT violation, 
Lorentz breaking is involved as well, 
so that boost- and rotation-dependent effects 
can be searched for.

{\bf Low-energy precision tests with neutrons.}
Another particle that
offers excellent possibilities for ultrahigh-sensitivity measurements 
is the neutron. 
Experimental difficulties that
would arise from its instability 
can be avoided by 
performing measurements with bound neutrons. 
Such measurements have placed limits down to $10^{-33}\,$GeV
on certain Lorentz- and CPT-violating neutron coefficients 
of the mSME~\cite{randompnexpt}.
A drawback of tests with bound neutrons is that
the primary source for theoretical errors 
is due to nuclear modeling. 
Experiments with free neutrons are therefore of interest as well. 
The first such experiment consisted of 
Larmor-frequency measurements with ultracold neutrons 
by the nEDM collaboration.
This test has placed the bound of $10^{-29}\,$GeV 
on Lorentz- and CPT-violation of the neutron~\cite{randompnexpt}.
Although of less sensitivity,
this experimental constraint provides a complementary and much cleaner measurements.

\section{Summary}
\label{sum}

At the present time, 
no convincing experimental evidence 
for Lorentz or CPT breakdown exists.
Nevertheless,
various approaches to more fundamental physics 
(e.g., quantum-gravity models)
contain mechanisms 
for generating feeble violations of Lorentz, CPT, and translational invariance. 
In this talk, 
a brief survey of the
motivations, 
theoretical ideas, 
and experimental efforts 
in the field of spacetime-symmetry tests
has been presented.

We have explained that 
a quantum description of the dynamics of spacetime 
is likely to require new spacetime concepts:
the smooth-manifold picture 
may have to be abandoned at Planck-size distance scales. 
Among the many possible observational signals
for such a quantum spacetime,
symmetry considerations are particularly promising
for two reasons.
First, 
symmetries are perhaps the only feature
of a putative quantum nature of spacetime 
amenable to Planck-precision tests 
at the present time.
Second, 
theoretical models can accommodate 
departures from spacetime symmetries,
which we have exemplified by
spontaneous symmetry breakdown in string field theory,
varying scalars,
and non-commutative geometry.

At energies that
can currently be reached in experiments,
general Lorentz- and CPT-violating effects
can be described by an effective field theory 
known as the SME. 
This framework incorporates practically all of established physics
(i.e., the Standard Model and general relativity), 
so that the Lorentz- and CPT-violating properties 
of essentially all physical systems can be studied,
at least in principle.
The SME coefficients for Lorentz and CPT breakdown  
are given by 
externally prescribed non-dynamical background vectors and tensors 
that are presumed to be caused 
by underlying physics.

Spacetime symmetries provide the basis for a wide variety of physical effects.
For this reason, 
Lorentz and CPT tests can be performed in a broad range 
of physical systems. 
This fact, 
together with the availability of the modern SME test framework 
and encouraging motivations for Lorentz and CPT violations
has resulted in the recent rise in experimental interest in the field.
We have reviewed 
a representative sample of experimental efforts 
along these lines 
including 
dispersion-relation analyses, 
spectropolarimetry of cosmological sources, 
and low-energy ultrahigh-precision laboratory studies.

A number of important open questions remain in this subject.
They are of foundational, of phenomenological, and of observational nature; 
they provide ample ground for future research in the field of spacetime-symmetry physics.

\section*{Acknowledgments}
The author wishes to thank Klaus Kirch 
for the invitation to this stimulating meeting.
This work was supported in part 
by CONACyT under Grant No.\ 55310  
and under the Red FAE program
as well as by the 
Funda\c{c}\~ao para a Ci\^encia e a Tecnologia
under Grant No.\ CERN/FP/109351/2009.








\end{document}